\documentclass[fleqn,pra]{revtex4-2}

\usepackage{amsmath,graphicx}
\usepackage{verbatim}

\begin{document}

\title{Forbidden "ortho"--"para" electric dipole transitions in H$_2^+$ ion.}

\author{Vladimir I. Korobov}
\affiliation{Bogoliubov Laboratory of Theoretical Physics, Joint Institute for Nuclear Research, Dubna 141980, Russia}
\author{D. Bakalov}%
\affiliation{Institute for Nuclear Research and Nuclear Energy, Bulgarian Academy of Sciences,\\
    blvd. Tsarigradsko ch. 72, Sofia 1142, Bulgaria}

\begin{abstract}
We present the first systematic calculation of the electric dipole forbidden transitions in the homonuclear molecular ion H$_2^+$. We get that the transition rate from the ground "ortho" $(v\!=\!0,N\!=\!1,J\!=\!1/2)$ state to the ground "para" $(v\!=\!0,N\!=\!0,J\!=\!1/2)$ state is $4.9\!\times\!10^{-14}$ s$^{-1}$ that corresponds to the lifetime of $6.4\!\times\!10^5$ yr. The Einstein coefficient $A_{nn'}$ for the $(v\!=\!1,N\!=\!1,J\!=\!1/2)\!\to\!(v\!=\!0,N\!=\!0,J\!=\!1/2)$ transition is equal to $0.12\!\times\!10^{-9}$ s$^{-1}$, which is of "comparable" order with the values for quadrupole transitions in H$_2^+$. It gives hope that the "ortho"--"para" transitions in H$_2^+$ can be induced by the laser spectroscopy.
\end{abstract}

\maketitle

\section{Introduction}

Forbidden E1 transitions between "ortho" (triplet in the proton spins, $I=1$) and "para" (singlet, $I=0$) states in the hydrogen molecular ion H$_2^+$ are of significant interest. So far, only transitions between near-threshold high vibrational $v\sim18\!-\!19$ states were studied numerically \cite{Bunker00} and experimentally \cite{Moss03}, where it was discovered that due to strong $g/u$ state mixing (\textit{gerade/ungerade} electronic symmetry states) near threshold the transition rate is high enough for spectroscopic measurements. For example, the rate of the $(v=19,N=1)\to(v=18,N=0)$ transition is $9.6\times10^{-6}$ s$^{-1}$.

In the case of hydrogen molecule H$_2$, Wigner gave a qualitative estimate for the E1 radiative decay rate from  "ortho" to "para" hydrogen as $R_{\rm est}=10^{-14}~\mbox{s}^{-1}$ (see \cite{Dodelson}). First calculations by Raich and Good \cite{Raich64} showed that the decay rate of the $X^1\Sigma^+_g(N\!=\!1)$ lowest vibrational "ortho" state is about $\sim\!2\!\times\!10^{-13}$ yr$^{-1}$, many orders of magnitude smaller than the nominal rate due to strong cancellations. Dodelson \cite{Dodelson} reanalyzed the calculations of Raich and Good using a fully quantum electrodynamical approach, introduced by Feinberg and Sucher \cite{FeinbergSucher} for calculations in the helium atom, and derived additional terms, which change the previous result by about 20\%. Later Pachucki and Komasa \cite{Pachucki08} extended the calculations by taking into account both E1 and M2 transitions, they've got the E1 decay rate of the $(v\!=\!0,N\!=\!1,J\!=\!1)$ "ortho" state equal to $1.68\!\times\!10^{-13}$ yr$^{-1}$ (or $5.3\!\times\!10^{-21}$ s$^{-1}$). The radiative lifetimes of the $1s\sigma_g$ $(v\!=\!0,N\!=\!1,J\!=\!1/2)$ "ortho" state of the H$_2^+$ molecular ion have been only roughly estimated in \cite{Bunker00}. Our calculation shows that this estimate was incorrect.

In the present work we intend to perfom calculations of the E1 forbidden transitions for the hydrogen molecular ion H$_2^+$ at low $v$ and $N$, which takes into account all relativistic corrections of order $(m_e/m_p)(Z\alpha)^2$ to the transition amplitude. We intend to show that in the case of the E1 decay, transition rates are much higher than for the neutral molecule and are in agreement with Wigner's estimate for this quantity.


In what follows we assume atomic units: $\hbar=|e|=m_e=1$.

\section{The Hamiltonian}

In this work we adopt the following notation: $v$ is the vibrational quantum number, $N$ is the total orbital angular momentum of the nonrelativistic wave-function. The spin part is described by the spin operators of two protons, $\mathbf{I}_1$, $\mathbf{I}_2$, and the spin of an electron, $\mathbf{s}_e$, $\mathbf{I}=\mathbf{I}_1+\mathbf{I}_2$ is the total nuclear spin, $\mathbf{F}=\mathbf{I}+\mathbf{s}_e$ is the total spin of the H$_2^+$ ion, and $\mathbf{J}=\mathbf{F}+\mathbf{N}$ is the total angular momentum. Thus the ground "para" state is denoted $(v=0,N=0,I=0,F=1/2,J=1/2)$. The excited rotational "ortho" state of the ground vibrational level $(v=0,N=1)$ has two spin states: $F\!=\!1/2$ and $F\!=\!3/2$. We will call the first state as the ground "ortho" state. While the triplet states with the total spin $F\!=\!3/2$ are not coupled to the subspace of the singlet nuclear spin states, at least within the first order corrections to the wave function considered in this work. For this reason we will ignore the $F\!=\!3/2$ states in our studies and will use the shorter notation for description of a state: $(v,N,J)$.

The Hamiltonian for an ion interacting with the radiation electromagnetic field, which also includes the leading order Breit-Pauli corrections, is expressed as follows:
\begin{equation}\label{Hamiltonian}
H = H_0 + H_{BP} + H_I,
\end{equation}
where $H_0$ is the zero-order nonrelativistic Hamiltonian of the particles, $H_{BP}$ is the Breit-Pauli Hamiltonian, and $H_I$ is the part which describes interaction of the ion with the electromagnetic field.

The nonrelativistic Hamiltonian in the center of mass frame may be written as:
\begin{equation}\label{H_0}
H_0 
 = \frac{\mathbf{p}_1^2}{2M}+\frac{\mathbf{p}_2^2}{2M}+\frac{\mathbf{p}_e^2}{2m_e}
      -\frac{Z}{r_1}-\frac{Z}{r_2}+\frac{Z^2}{R},
\end{equation}
where $\mathbf{r}_i = \mathbf{r}_e\!-\!\mathbf{R}_i$ and $\mathbf{R} = \mathbf{R}_2\!-\!\mathbf{R}_1$ are electron coordinates relative to nuclei and internuclear position vector in the molecular coordinate notations, $(\mathbf{r}_e,\mathbf{R}_1,\mathbf{R}_2)$ and $(\mathbf{p}_e,\mathbf{p}_1,\mathbf{p}_2)$ are the position vectors and momenta of particles in the center of mass frame, $M=m_p$ is the proton mass, and $Z=1$ is the proton charge. According to the tradition accepted in the theory of the light atoms we will use $Z$ for the proton charge in order to distinguish between two scales: $\alpha$, fine structure constant, and $v/c\approx Z\alpha$, mean particle velocity in a bound system.

The Breit-Pauli Hamiltonian $H_{BP}$ is the leading order relativistic correction (see, for example, \cite{BS}). Here we need only the spin-spin and spin-orbit interactions responsible for the $g/u$ mixing; we will follow the notations of \cite{HFS06}.
The spin-spin and spin-orbit interactions are expressed, respectively:
\begin{equation}
H_{ss} = \alpha^2\frac{(1\!+\!\kappa_e)(1\!+\!\kappa_p)}{m_em_p}\frac{8\pi}{3}
         \left[
            (\mathbf{s}_e\cdot\mathbf{I}_1)\delta(\mathbf{r}_1)
            +(\mathbf{s}_e\cdot\mathbf{I}_2)\delta(\mathbf{r}_2)
         \right].
\end{equation}
\begin{equation}
\begin{array}{@{}l}\displaystyle
H_{so} = \alpha^2\Biggl\{
            \frac{1\!+\!2\kappa_e}{2m_e^2}
            \left[
               \frac{[\mathbf{r}_1\times\mathbf{p}_e]}{r_1^3}
               +\frac{[\mathbf{r}_2\times\mathbf{p}_e]}{r_2^3}
            \right]\mathbf{s}_e
            -\frac{1\!+\!\kappa_e}{m_em_p}
            \left[
               \frac{[\mathbf{r}_1\times\mathbf{p}_1]}{r_1^3}
               +\frac{[\mathbf{r}_2\times\mathbf{p}_2]}{r_2^3}
            \right]\mathbf{s}_e
\\[3mm]\displaystyle\hspace{15mm}
            -\frac{1\!+\!2\kappa_p}{2m_p^2}
            \left[
               \frac{[\mathbf{r}_1\times\mathbf{p}_1]}{r_1^3}\mathbf{I}_1
               +\frac{[\mathbf{r}_2\times\mathbf{p}_2]}{r_2^3}\mathbf{I}_2
            \right]
            +\frac{1\!+\!\kappa_p}{m_em_p}
            \left[
               \frac{[\mathbf{r}_1\times\mathbf{p}_e]}{r_1^3}\mathbf{I}_1
               +\frac{[\mathbf{r}_2\times\mathbf{p}_e]}{r_2^3}\mathbf{I}_2
            \right]
         \Biggr\}.
\end{array}
\end{equation}

The operators which connect ortho and para states are
\begin{equation}
\begin{array}{@{}l}\displaystyle
H_{ss}^- = \alpha^2\frac{(1\!+\!\kappa_e)(1\!+\!\kappa_p)}{m_em_p}\frac{4\pi}{3}
         \Bigl\{
            (\mathbf{s}_e\cdot\mathbf{I}_-)\bigl[\delta(\mathbf{r}_1)\!-\!\delta(\mathbf{r}_2)\bigr]
         \Bigr\},
\\[4mm]\displaystyle
H_{so}^- = \alpha^2\Biggl\{
            \frac{1\!+\!\kappa_p}{2m_em_p}
            \left(
               \frac{[\mathbf{r}_1\times\mathbf{p}_e]}{r_1^3}
               -\frac{[\mathbf{r}_2\times\mathbf{p}_e]}{r_2^3}
            \right)\mathbf{I}_-
            -\frac{1\!+\!2\kappa_p}{4m_p^2}
            \left(
               \frac{[\mathbf{r}_1\times\mathbf{p}_1]}{r_1^3}
               -\frac{[\mathbf{r}_2\times\mathbf{p}_2]}{r_2^3}
            \right)\mathbf{I}_-
         \Biggr\}
\end{array}
\end{equation}
where $\mathbf{I}_1$ and $\mathbf{I}_2$ are the spin operators of the two protons, $\mathbf{I}_- = (\mathbf{I}_1-\mathbf{I}_2)$, $\kappa_e$ and $\kappa_p$ are the magnetic moment anomaly of an electron and proton, respectively, while $\mu_p=1\!+\!\kappa_p$ is the magnetic moment of the proton in nuclear magnetons.

In our consideration we use the NRQED formalism \cite{Lepage86,Kinoshita96}. For our needs only the leading order terms of the Lagrangian \cite{Kinoshita96}, Eq.~(19), are needed both for an electron and proton. We use the Coulomb gauge, and the one-photon interaction Hamiltonian (in the center of mass frame) may be expressed as:
\begin{equation}\label{H_I}
\begin{array}{@{}l}\displaystyle
H_I = \sum_a Z_a\alpha\frac{\mathbf{p}_a}{m_a}\mathbf{A}_r
   +\sum_a \frac{Z_a\alpha(1\!+\!\kappa_a)}{2m_a}\boldsymbol{\sigma}_a\mathbf{B}_r
   -\sum_a \frac{Z_a\alpha(1\!+\!2\kappa_a)}{8m_a^2}\:
       \boldsymbol{\sigma}_a
       \Bigl(\mathbf{p}_a\!\times\!\mathbf{E}_{\perp}-\mathbf{E}_{\perp}\!\times\!\mathbf{p}_a\Bigr)
\\[3mm]\displaystyle\hspace{8mm}
   +\sum_a \frac{Z_a^2\alpha^2}{2m_a}\mathbf{A}\!\cdot\!\mathbf{A}_r
   -\sum_a \frac{Z_a^2\alpha^2(1\!+\!2\kappa_a)}{4m_a^2}\:
       \boldsymbol{\sigma}_a
       \Bigl(\mathbf{A}_r\!\times\!\mathbf{E}_{\parallel}\Bigr)
   +\sum_a \frac{Z_a^2\alpha^2(1\!+\!2\kappa_a)}{4m_a^2}\:
       \boldsymbol{\sigma}_a
       \Bigl(\mathbf{E}_{\perp}\!\times\!\mathbf{A}\Bigr)+\dots,
\end{array}
\end{equation}
we use $\mathbf{A}_r$, $\mathbf{B}_r$, and $\mathbf{E}_{\perp}$ to denote operators of the external electromagnetic radiation field. The transverse fields $\mathbf{E}_{\perp}$ and $\mathbf{B_r}$ depend on $\mathbf{A}_r$ as
$\mathbf{E}_{\perp} = -\frac{1}{c}\frac{\partial}{\partial t}\mathbf{A}_r(\mathbf{r},t)$ and
$\mathbf{B}_r = \boldsymbol{\nabla}\!\times\!\mathbf{A}_r(\mathbf{r},t)$.

Operators $\mathbf{E}_{\parallel}$ and $\mathbf{A}$ are the electric-field strength and the magnetic-field potential, which are induced by the particles constituting the molecular ion. For our derivations we need only the expression for the magnetic-field potential produced by the magnetic moment of the nuclei (corresponds to the $\mathbf{A}\!\cdot\!\mathbf{A}_r$ term of the electron line in (\ref{H_I})):
\[
\mathbf{A} = -\frac{Z\alpha(1\!+\!\kappa_p)}{m_p}
   \left(
      \frac{\left[\mathbf{r}_1\!\times\!\mathbf{I}_1\right]}{r_1^3}
      +\frac{\left[\mathbf{r}_2\!\times\!\mathbf{I}_2\right]}{r_2^3}
   \right),
\]
which results in correction to the nonrelativistic current $\mathbf{J}^{(0)}=\sum_a Z_a\alpha\mathbf{p}_a/m_a$:
\begin{equation}\label{Jcor}
\boldsymbol{\delta}\mathbf{J} =
   -\frac{Z\alpha^3 (1\!+\!\kappa_p)}{2m_em_p}
   \left\{
      \frac{\left[\mathbf{r}_1\!\times\!\mathbf{I}_1\right]}{r_1^3}
      +\frac{\left[\mathbf{r}_2\!\times\!\mathbf{I}_2\right]}{r_2^3}
   \right\}.
\end{equation}

The other terms in Eq.~(\ref{H_I}) produce corrections to the electric current, which either do not contain the proton spin operator, or are of higher order in $m_e/m_p$ or $Z\alpha$ expansion and thus will be neglected.

\section{Transition amplitudes}

\begin{table}[b]
\begin{center}
\begin{tabular}{c@{\hspace{6mm}}r@{\hspace{6mm}}r}
\hline\hline
transition  & this work & \cite{Bunker00}~~ \\
\hline
$(19,0)\!\to\!(18,1)$ &  11.5 & 11.5 \\
$(19,0)\!\to\!(17,1)$ & 161.4 & 163.7 \\
\hline\hline
\end{tabular}
\end{center}
\caption{Comparison with calculations by Bunker and Moss \cite{Bunker00}. Averaged transition rates $\bar{A}_{nn'}$ (in $10^{-6}$ s$^{-1}$). 
}\label{T:comp}
\end{table}

The transition amplitude for the forbidden E1 transitions is expressed
\begin{equation}
\begin{array}{@{}l}\displaystyle
\mathcal{T}_{E1}^i =
   \left\langle
      \psi_{n}\left|J^i\right|\psi_{n'}
   \right\rangle +
   \left\langle
      \psi_n\left|J^{(0)i}\right|\psi_{n'}^{(1)}
   \right\rangle +
   \left\langle
      \psi_n^{(1)}\left|J^{(0)i}\right|\psi_{n'}
   \right\rangle
 = \left\langle
      \psi_{n}\left|J^{(0)i}\right|\psi_{n'}
   \right\rangle
   +\left\langle
      \psi_{n}\left|\delta J^{i}\right|\psi_{n'}
   \right\rangle +
\\[3mm]\hspace{12mm}\displaystyle
   +\left\langle
      \psi_n\left|J^{(0)i}Q(E_{n'}\!-\!H_0)^{-1}QH_{\text{gu}}\right|\psi_{n'}
   \right\rangle +
   \left\langle
      \psi_n\left|H_{\text{gu}}Q(E_n\!-\!H_0)^{-1}QJ^{(0)i}\right|\psi_{n'}
   \right\rangle,
\end{array}
\end{equation}
where $\mathbf{J}$ is the electric current operator of the system of bound particles (ion), $\psi_{n}$ and $\psi_{n'}$ are the nonrelativistic solutions of the Schr\"odinger equation, $\psi_{n}^{(1)}$ and $\psi_{n'}^{(1)}$ are the first order relativistic corrections to the wave functions, $Q$ is the projection operator on a subspace orthogonal to $\psi_n$ (or $\psi_{n'}$), and the operator
\[ H_{\text{gu}}=H_{so}^{(-)}+H_{ss}^{(-)} \]
is the $g/u$ symmetry breaking part of the Breit-Pauli Hamiltonian.

Transition amplitude in the length form may be obtained, using relation: $[H_0,\mathbf{r}_a]=-i\,\mathbf{p}_a/m_a$, or $\mathbf{J}^{(0)}=i\alpha[H_0,\mathbf{d}]$. Then one gets
\begin{equation}\label{Eq:T}
\begin{array}{@{}l}\displaystyle
\mathcal{T}_{E1}^i =
   \left\langle
      \psi_{n}\left|\mathcal{D}^i\right|\psi_{n'}
   \right\rangle
   +i\,\alpha w_{nn'}\left\langle
      \psi_n\left|d^iQ(E_{n'}\!-\!H_0)^{-1}QH_{\text{gu}}\right|\psi_{n'}
   \right\rangle
\\[3mm]\displaystyle\hspace{32mm}
   +i\,\alpha w_{nn'}\left\langle
      \psi_n\left|H_{\text{gu}}Q(E_n\!-\!H_0)^{-1}Qd^i\right|\psi_{n'}
   \right\rangle
\end{array}
\end{equation}
where $d^i = \sum_a Z_a r_a^i$ is the dipole operator in the length form, and
\[
\mathcal{D}^i =
   i\,\alpha w_{nn'}d^i-i\alpha\left[d^i,H_{\rm gu}\right]+\delta J^{i}.
\]
Here the commutator is
\begin{equation}
i\left[\mathbf{d},H_{\rm gu}\right] = -\frac{\alpha^2}{2}\left(1+\frac{1}{m_p}\right)\frac{(1\!+\!\kappa_p)}{m_em_p}
   \left(\frac{[\mathbf{r}_1\!\times\!\mathbf{I}_1]}{r_1^3}+\frac{[\mathbf{r}_2\!\times\!\mathbf{I}_2]}{r^3_2}\right)
\end{equation}
and it cancels out the electric current correction term, Eq.~(\ref{Jcor}), in the leading order in $m_e/m_p$.

\section{Results and discussion}

Numerical calculations were based on the "exponential" variational expansion \cite{Korobov00}. Particularly, the nonrelativistic solutions $\psi_n$ and the first order corrections to the wave functions, $\psi_n^{(1)}$, were calculated using this expansion method. By averaging Eq.~(\ref{Eq:T}), we obtained the transition amplitudes $T_{nn'}$ and, eventually, the Einstein coefficients $A_{nn'}$ for the spontaneous emission of a photon from the state $n=(v,N,J)$ to the state $n'=(v',N',J')$.

In Table \ref{T:comp} our results are compared with the results of Bunker and Moss \cite{Bunker00}. In order to make this comparison, the transition rates are averaged over $J'$,
\begin{equation}\label{Eq:average}
\bar{A}_{nn'}=\bar{A}_{N=0,N'=1}=\frac{2}{3}A_{J=\frac{1}{2},J'=\frac{3}{2}}+\frac{1}{3}A_{J=\frac{1}{2},J'=\frac{1}{2}}.
\end{equation}
In \cite{Bunker00} only the spin-spin interaction has been taken into account as a source of the $g/u$ breaking. That is justified for the weakly bound states where the spin-spin coupling becomes dominant due to proximity of the $2p\sigma_u$ states.

Tables \ref{T:0-1} and \ref{T:1-2} present the results of our numerical calculations for $N\!=\!0\!\to\!1$ and $N\!=\!1\!\to\!2$ transitions, respectively.
As is seen transition rates for $\Delta v=1$ transitions are of "comparable" order of magnitude with the rate of quadrupole transitions in H$_2^+$ \cite{Baye12,H2+_quadrupole}. For example, the E2 transition: $(v\!=\!0,N\!=\!0)\to(0,2)$, has $A_{nn'}=9.7\!\times\!10^{-12}$ s$^{-1}$, and the E2 transition: $(v\!=\!0,N\!=\!0)\to(2,2)$, has $A_{nn'}=32.\!\times\!10^{-9}$ s$^{-1}$ \cite{H2+_quadrupole}. This fact allows to hope that the "ortho"--"para" transitions may be studied in laser spectroscopic experiments.

It is interesting to note that the contribution of spin-orbit interactions to the decay rate of low $v$ states is about 5-10\%. It turns out that a simple approach with only spin-spin coupling should give fairly good results. In this case, only the reduced matrix element of the transition $N\to N'$ can be calculated, and the spin part of the amplitude can be expressed using Wigner's 6-$j$ symbols, as for allowed transitions:
\[
\left\langle vFNJ\|\mathbf{d}\|v'FN'J'\right\rangle =
   (-1)^{J+F+N'+1}\sqrt{(2J\!+\!1)(2J'\!+\!1)}
   \left\{\begin{matrix}
      N & 1 & N' \\ N' & F & J
   \end{matrix}\right\}
   \left\langle vN\|\mathbf{d}\|v'N'\right\rangle\,.
\]

The terms, which were derived in the Dodelson work (see Eq.~(32), \cite{Dodelson}) and were essential for the molecule, in case of H$_2^+$ ion have contributions of the same magnitude as for the H$_2$. Thus they have a negligible effect on the transition rate in the H$_2^+$ molecular ion.

\begin{table}
\begin{center}
\begin{tabular}{c@{\hspace{6mm}}c@{\hspace{6mm}}c}
\hline\hline
$v\!\to\!v'$ & $J=\frac{1}{2}\!\to\!J'=\frac{1}{2}$ & $J=\frac{1}{2}\!\to\!J'=\frac{3}{2}$ \\
\hline
$0\!\to\!0$ & 0.0558 & 0.0494 \\
$0\!\to\!1$ &  123.7 &  128.3 \\
$0\!\to\!2$ & 0.1170 & 0.0708 \\
$0\!\to\!3$ & 0.0429 & 0.0554 \\
$0\!\to\!4$ & 0.0155 & 0.0183 \\
\hline
$1\!\to\!0$ &  238.7 &  122.7 \\
$1\!\to\!1$ & 0.0646 & 0.0585 \\
$1\!\to\!2$ &  274.3 &  283.1 \\
$1\!\to\!3$ & 0.8526 & 0.6433 \\
$1\!\to\!4$ & 0.1131 & 0.1510 \\
\hline\hline
\end{tabular}
\end{center}
\caption{The Einstein coefficient, $A_{nn'}$, (in $10^{-12}$ s$^{-1}$) for transitions between "ortho" and "para" states, $N=0\!\to\!N=1$}\label{T:0-1}
\end{table}

\begin{table}
\begin{center}
\begin{tabular}{c@{\hspace{6mm}}c@{\hspace{6mm}}c@{\hspace{6mm}}c}
\hline\hline
$v\!\to\!v'$ & $J=\frac{3}{2}\!\to\!J'=\frac{5}{2}$ & $J=\frac{3}{2}\!\to\!J'=\frac{3}{2}$ & $J=\frac{1}{2}\!\to\!J'=\frac{3}{2}$ \\
\hline
$0\!\to\!0$ & 0.5229 & 0.0788 & 0.4186 \\
$0\!\to\!1$ &  151.8 &  26.15 &  128.3 \\
$0\!\to\!2$ & 0.0817 & 0.0070 & 0.0527 \\
$0\!\to\!3$ & 0.0561 & 0.0121 & 0.0523 \\
$0\!\to\!4$ & 0.0195 & 0.0039 & 0.0176 \\
\hline
$1\!\to\!0$ &  211.3 &  23.96 &  236.6 \\
$1\!\to\!1$ & 0.6076 & 0.0932 & 0.4899 \\
$1\!\to\!2$ &  337.1 &  57.82 &  284.3 \\
$1\!\to\!3$ & 0.7216 & 0.0865 & 0.5265 \\
$1\!\to\!4$ & 0.1511 & 0.0335 & 0.1426  \\
\hline\hline
\end{tabular}
\end{center}
\caption{The Einstein coefficient, $A_{nn'}$, (in $10^{-12}$ s$^{-1}$) for transitions between "ortho" and "para" states, $N=1\!\to\!N=2$.}\label{T:1-2}
\end{table}

\section*{Acknowledgements}

The authors acknowledge support from the European Research Council (ERC) under the European Union’s Horizon 2020 research and innovation programme (Grant Agreement No.~786306 “PREMOL” ERC-2017-AdG).

\end{document}